\documentclass[12pt]{article}
\usepackage{latexsym}
\usepackage{amssymb}
\usepackage{cite}
\usepackage{hyperref}
\usepackage{graphicx}
\title{\LARGE Brachistochrone and Sliding with Friction}
\author{\large A.V.Kurilin{\footnote{E-mail address:
kurilin@mail.ru}}\\ Moscow Technical University of Communications and Informatics (MTUCI) \footnote{on leave from Moscow Technological Institute}\\ 
\\}
\date{March 24, 2023}

\begin{document}
\begin{titlepage}
\maketitle \thispagestyle{empty}

\begin{abstract}{Motions of a material point along a set of parabolas are studied, taking into account the forces of Coulomb friction. The obtained results are compared with similar motions along the cycloid. The analysis is carried out using numerical calculations in the Mathcad program.}
\end{abstract}
\end{titlepage}

The problems of classical mechanics underlie modern physics and serve as a kind of standard in the development of new methods for solving tasks in modern natural science. One of such striking examples in the history of science is the well-known problem of Johann Bernoulli about the brachistochrone \cite{Tikhomirov}, formulated in 1696. Many of the greatest scientists, whose names are now cited as the classics of science, were engaged in solving the problem of the brachistochrone \cite{Courant}.
Let us recall its essence. Given two points $A$ and $B$ lying in a vertical plane, which are separated by some vertical distance $H$ and a distance $L$ horizontally. The question is what is the trajectory of a point moving only under the influence of gravity, which starts moving from 
$A(0;H)$ and reaches the point $B(L;0)$ in the minimal time? It is known that five different solutions to this problem were obtained, which agreed in the fact that the brachistochrone, the curve of the fastest descent, is a cycloid given by the following parametric equations:
\begin{equation}
\label{cycloid-1}
\left\{
{\begin{array}{l}
x(\tau)=C_{0}(\tau-\sin(\tau)),\\
y(\tau)=H-C_{0}(1-\cos(\tau)).\\
 \end{array}} \right. 
\end{equation}
The parameter $C_{0}$ in the equation (\ref{cycloid-1}) for the cycloid is related to the Cartesian coordinates of the points $x_{A}=0, y_{A}=H$   and $ x_{B}=L, y_{B}=0$ through the relations:
\begin{equation}
\label{cycloid-2}
\left\{
{\begin{array}{l}
L=C_{0}(\tau_{0}-\sin(\tau_{0})),\\
H=C_{0}(1-\cos(\tau_{0})),\\
\end{array}} \right.
 \end{equation}
where $\tau_{0}$ is the limiting value of the parameter  included in the cycloid equation (\ref{cycloid-1}): $0\le \tau \le \tau_{0}.$
In recent years, the problem of the brachistochrone has again attracted new attention of many researchers in attempts to find its reasonable generalizations taking into account the influence of resistance forces, the finite dimensions of a moving body, relativistic effects, etc.\cite{Sumbatov-MIPT}. It is worth to single out the works devoted to investigations of the trajectory of brachistochrone when the force of dry Coulomb friction is taken into account \cite{AJM-1975,Lipp,Hayen,Golubev-2010,Golubev-2012,Sumbatov-2017}. Several results were obtained describing the required curve of the fastest descent in an implicit, rather complicated form, which is extremely inconvenient for numerical analysis. In this regard, it would be interesting to calculate the parameters of such movements by computer methods and track how the particle velocity, its coordinates and the total descent time changes depending on the shape of the trajectory and its curvature. In our previous work \cite{Kurilin-2016}, the parameters of free motion along smooth trajectories without friction were calculated. In this article some results of similar calculations taking into account the forces of Coulomb friction are presented and compared with the data from other studies.

\begin{figure}[h]
\noindent
\begin{center}
\includegraphics[width=130mm]{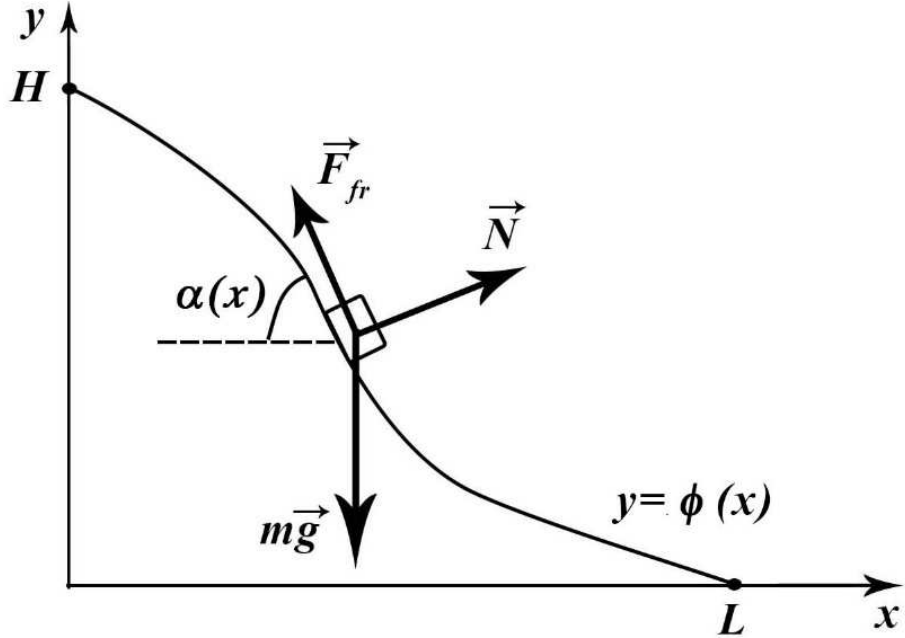}
\caption[]{Descent curve under the influence of gravity and dry friction.}
\label{Hill}
\end{center}
\end{figure}

\begin{figure}[h]
\begin{center}
\includegraphics[width=150mm]{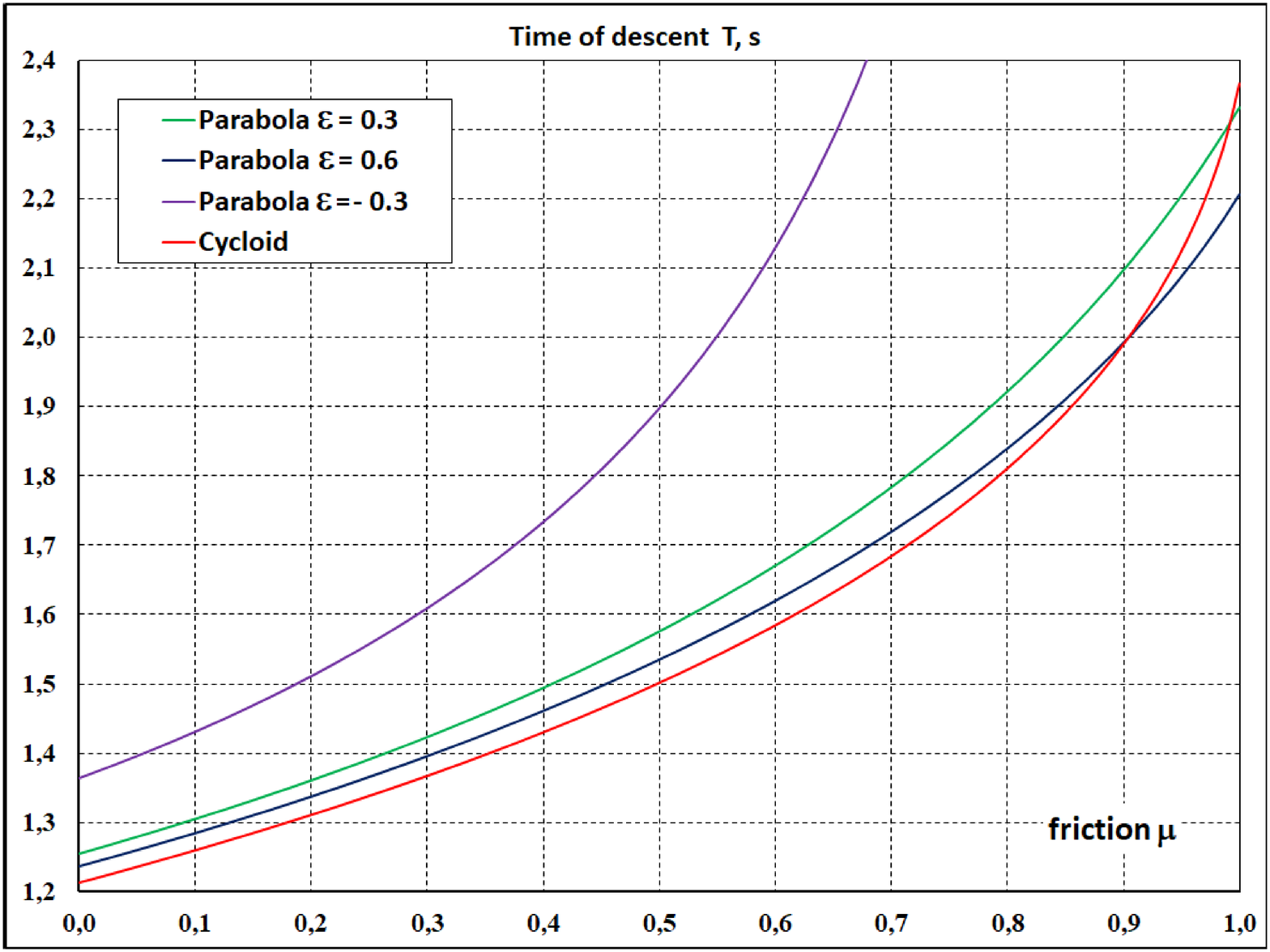}
\caption[]{Dependence of the descent time (\ref{TimeParabola}) on the coefficient of friction $\mu$.}
\label{Time-Parabolas3}
\end{center}
\end{figure}

Let’s consider a material point of mass $m$ sliding down along some curved line given by the explicit function: $y=\phi(x)$. The classical equations of Newton's dynamics can be written in the system (\ref{Newton-1}) and  relate the inclination angle of the trajectory at a given point $\tan\alpha(x)=-\phi^{\prime}(x)$  (see Fig.\ref{Hill}) with tangent part of acceleration  $a_{\tau} $, which determines changes in the modulus of the velocity vector $v=\sqrt{\dot x^2+\dot y^2}=\dot x\sqrt{1+\phi^{\prime}(x)^2} $. The force of dry friction $F_{fr}=\mu N$ depends on the friction coefficient $\mu$ and the force of the normal reaction $N$.
\begin{equation}
\label{Newton-1}
\left\{
{\begin{array}{l}
mg \sin\alpha-\mu N=ma_{\tau}=m{dv \over dt},\\
N- mg\cos\alpha=ma_{n}=\frac{mv^2}{R},\\
\end{array}} \right.
 \end{equation} 	
The curvature radius $R$ of the trajectory at a given point $\left({x;\,\phi (x)} \right)$ 
as well as inclination angle $\alpha (x)$ can be 
expressed from the equation of the curve $y=\phi (x)$:
\begin{equation}
\label{angle-3}
\left\{ {\begin{array}{l}
 \sin \alpha \left( x\right)=\frac{-\phi^{\prime}(x)}{\sqrt {1+\phi^{\prime}(x)^2}}, 
\\ 
 \cos \alpha \left( x \right)=\frac{1}{\sqrt {1+\phi^{\prime}(x)^2}}, \\ 
 \end{array}} \right.\,\quad \,
 \frac{1}{R(x)}=\frac{\phi^{\prime\prime}(x)}{\left[ {1+\phi^{\prime}(x)^2} \right]^{3/2}}\,.
\end{equation}
Substituting eqn.(\ref{angle-3}) in (\ref{Newton-1}) we arrive to the following differential equation describing changes of the horizontal coordinate $x=x(t)$ over time:
 \begin{equation}
\label{DiffeqnX}
\ddot {x}(t)=-\frac{\phi^{\prime}(x)+\mu }{1+\phi^{\prime}(x)^2}
\left[ {g+\phi^{\prime\prime}(x)\dot {x}^2(t)} \right].
\end{equation}
Here, as usual, the prime at the sign of the function indicates its derivatives with respect to the $x$ variable, while the dot denotes to time derivatives. The order of this differential equation can be reduced by introducing an auxiliary function $W(x)$ which is equal to the square of the velocity of horizontal movement:
\begin{equation}
\label{Wfunction}
W(x)=\dot {x}^2(t)=\left( {\frac{dx}{dt}}\right)^2.
\end{equation}
Thus we proceed to the first order differentiation with respect to the variable $x$:
 \begin{equation}
\label{WdiffEqn}
\frac{1}{2}\frac{dW(x)}{dx}=-\frac{\phi^{\prime}(x)+\mu }{1+\phi^{\prime}(x)^2}
\left[ {g+\phi^{\prime\prime}(x)W(x)} \right].
\end{equation}
The obtained linear inhomogeneous equation can be solved in quadratures by the method of variation of constants
\begin{equation}
\label{Weqn}
W(x)=\frac{K(x)}{1+\phi^{\prime}(x)^2}\exp \left[ {-2\mu \cdot \mbox{arctan}\left( \phi^{\prime}(x)\right)} \right].
\end{equation}
Taking into account the initial conditions, ${x(0)\!=\!0}, \, {\dot{x}(0)\!=\!0}$, the constant variation  function $K(x)$ can be written as:
\begin{equation}
\label{Kfunction}
K(x)=-2g\int\limits_0^x {\left[ {\mu +\phi^{\prime}(\xi )} \right]} 
\exp\left[ {2\mu \cdot \mbox{arctan}\left( {\phi^{\prime}(\xi )} \right)}\right]d\xi .
\end{equation}
The physical meaning of this formula is quite obvious. The speed of movement at a given point on the trajectory is determined locally by the action of the gravity and friction forces, as well as by the non-local prehistory of movements, which takes into account accelerations in earlier times. The total descent time can now be found by integration along the horizontal coordinate:
\begin{equation}
\label{TimeWint}
T_0 =\int\limits_0^L {\frac{dx}{\dot {x}(t)}} =\int\limits_0^L 
{\frac{dx}{\sqrt {W\left( x \right)} }}.
\end{equation}
As an example of the use of these formulas, it is proposed to compare the laws of motion along a parabola and a cycloid, taking into account the force of dry friction and using numerical methods for integrating expressions (\ref{Weqn})~--~(\ref{TimeWint}). The parabola equation that satisfies initial conditions of the problem, $\phi (0)\!=\!H,\, \phi (L)\!=\!0$, can be chosen with one free parameter $\varepsilon $, which is responsible for the curvature of the trajectory.
\begin{equation}
\label{parabolas}
\phi (x)=\frac{\kappa \varepsilon }{L}x^2-\kappa \left( {1+\varepsilon } 
\right)x+H,
\quad
\kappa =\frac{H}{L}.
\end{equation}
\begin{figure}[h]
\setlength{\unitlength}{1cm}
\begin{center}
\includegraphics[width=150mm]{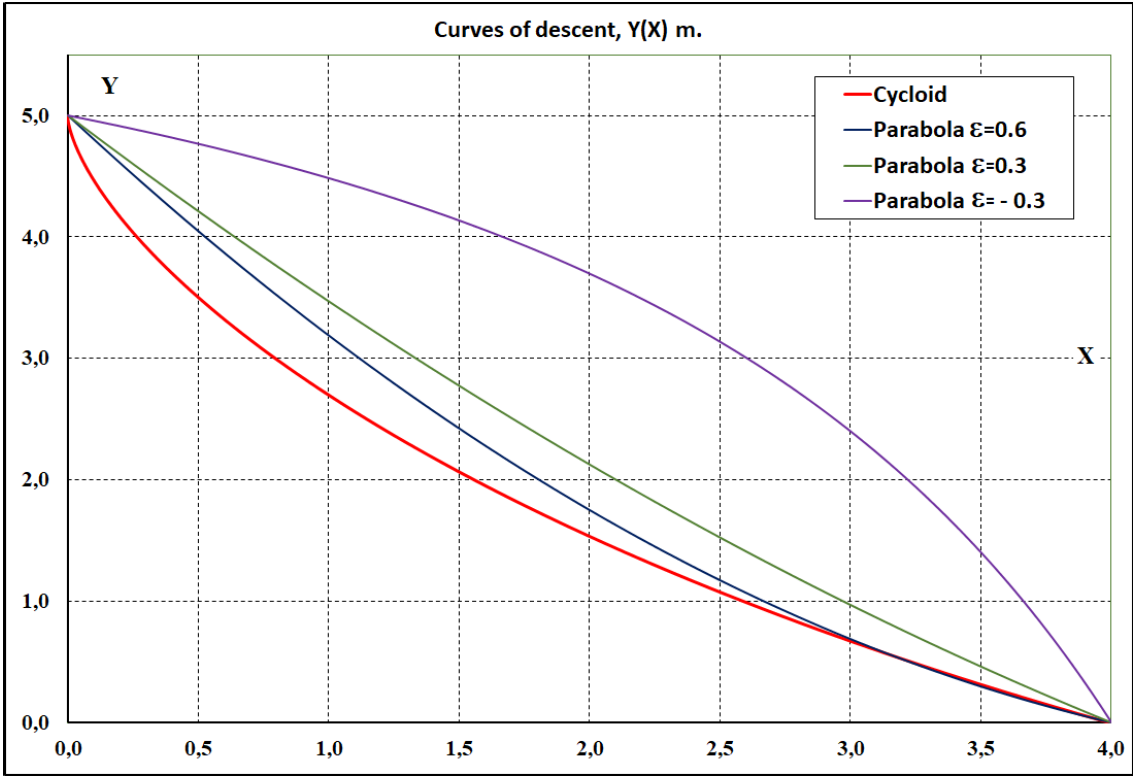}
\caption[]{The shape of the descent trajectories in the gravity field, on which the influence of the dry friction force was studied.}
\label{Curves}
\end{center}
\end{figure}
In the case, $\varepsilon >0$, we obtain a concave curve similar to the cycloid (\ref{cycloid-1}), which was already mentioned as the brachistochrone with zero sliding friction. Negative values, $\varepsilon <0$, correspond to convex trajectories, for which movements without an initial speed are possible only for limited values of the friction coefficient $\mu <\kappa (1+\varepsilon )$. In the limit $\varepsilon \to 0$ we arrive to sliding along inclined plane with a well-known descent time:
\begin{equation}
\label{PlaneTime}
T_1 =\sqrt {\frac{2L}{g}\left( {\frac{1+\kappa ^2}{\kappa -\mu }} \right)}.
\end{equation}
It is more convenient to describe the movement along a parabola in a parametric form, using the angle of inclination of the trajectory
$\alpha (x)=-\mbox{arctan}(\phi^{\prime}(x))$, varying from $\alpha _0$ to $\alpha _F$, where
\begin{equation}
\label{Angles-2}
\alpha _0 =\mbox{arctan}\left( {\kappa +\kappa \varepsilon} \right),
\quad
\alpha _F =\mbox{arctan}\left( {\kappa -\kappa \varepsilon} \right).
\end{equation}
Passing in formulas (\ref{Weqn}), (\ref{Kfunction}) to the parametric form of writing
\begin{equation}
\label{XparA}
x=\frac{L}{2\varepsilon }\left( {1+\varepsilon -\frac{1}{\kappa 
}\mbox{tan}\alpha } \right)
\end{equation}
we obtain the expression for the square of the horizontal speed (\ref{Wfunction})
\begin{equation}
\label{Wparabola}
W(\alpha )=\frac{gL}{\kappa \varepsilon }\cos^2\alpha e^{2\mu \alpha}\cdot FP\left( \alpha \right),
\end{equation}
which is written with the auxiliary integral function of the accelerating history:
\begin{equation}
\label{FParabola}
FP\left( \alpha \right)=\int\limits_\alpha ^{\alpha _0} 
{\left( {\mbox{tan}\beta -\mu } \right)\frac{e^{-2\mu \beta }}{\cos ^2\beta }d\beta}.
\end{equation}
The total time of descent along the parabolas (\ref{parabolas}) can now be represented as follows:
\begin{equation}
\label{TimeParabola}
T_0 =\frac{1}{2}\sqrt {\frac{L}{g\varepsilon \kappa }} \cdot 
\int\limits_{\alpha _F }^{\alpha _0 } {\frac{e^{-\mu \beta }}{\cos ^3\beta 
}\frac{d\beta }{\sqrt {FP(\beta )} }} .
\end{equation}
Numerical calculations in the Mathcad program using formulas (\ref{Wparabola}) - (\ref{TimeParabola}) were carried out for three different parabolas with different degrees of curvature $\varepsilon =-\,0.3;\,\,0.3;\,\,0.6$, shown in Figure \ref{Curves}. 
The following values were chosen as the initial parameters of the problem.
\begin{equation}
\label{LHData}
H=5\,m,\quad \,L=4\,m,\,\quad \,g=9.81\,\,ms^{-2}.
\end{equation}
The results of the calculations are depicted in Figure \ref{Time-Parabolas3}, which shows the descent time along the parabolas (\ref{parabolas}) depending on the friction coefficient $\mu$.
The limiting value of the friction coefficient at which downward movement is still possible is determined by the type of trajectory and the initial steepness. So for a convex trajectory $\varepsilon =-\,0.3$ with given parameters of problem (\ref{LHData}), this constraint is given by the initial inclination of the parabola $\mu <0.875$. For concave trajectories this constraint can be found only numerically. For example, when $\varepsilon 
=1.0$ the end point of the movement has zero steepness, so the last part of the descent will occur with a decreasing velocity. The limiting value of the friction coefficient in this case is determined by the fact that the modulus of the velocity vector at any point of the trajectory, except for the initial one, should not be equal to zero. Calculations show that this is possible for $\mu <0.8321$  with parabola $\varepsilon =1.0$, $\mu <1.0771$  for $\varepsilon =0.6$ and $\mu <1.2047$  for $\varepsilon =0.3$. Thus, it becomes clear that in contrast to the classical Bernoulli problem of brachistochrone, formulated for smooth curves, there is no universal curve capable of providing the minimum descent time for any values of the friction coefficient. At small values,  $\mu \ll 0.1$, the brachistochrone with dry friction will obviously resemble a cycloid, so it would be reasonable to calculate the time of movement along the cycloid, taking into account the friction force and with the same initial conditions. In this case, the parametric equation of the cycloid (\ref{cycloid-1}) can be rewritten in terms of the angle of inclination of the trajectory $\alpha (x)$  (\ref{angle-3}), using their interrelationship:
\begin{equation}
\label{Alpfa-Tau}
\alpha (x)=\alpha \left( {x(\tau )} \right)=\frac{\pi -\tau }{2}.
\end{equation}
The parameters of the cycloid (\ref{cycloid-1}) are calculated from the solution of the system of transcendental equations (\ref{cycloid-2}) and for the case (\ref{LHData}), are equal to:
\begin{equation}
\label{CycloidData}
C_0 =\mbox{3.40517104}\,m, \quad\,\tau _0 =\mbox{2.0582244}.
\end{equation}
The time of movement along the cycloid (\ref{cycloid-1}) with zero friction coefficient $\mu \to 0$ is \cite{Kurilin-2016}:
\begin{equation}
\label{CycloidTime0}
T_0 =\tau _0 \sqrt {\frac{C_0 }{g}} =1.213\,s.
\end{equation}
The function of the squared horizontal velocity (\ref{Wfunction}) - (\ref{Kfunction}) can now be expressed explicitly
\begin{equation}
\label{Wcycloid}
WC(\tau )=\frac{2g C_0 }{1+\mu ^2}\sin ^2\left( {\frac{\tau }{2}} 
\right)\cdot {\rm H}\left( \tau \right),
\end{equation}
where
\begin{equation}
\label{Fcycloid}
{\rm H}\left( \tau \right)=2 e^{-\mu \tau}-1-\left( {1-\mu ^2} 
\right) \cos \tau +2\mu \sin\tau-\mu^2.
\end{equation}
Thus it easy to calculate the time of descent along the cycloid (\ref{cycloid-1}) in a wide range of friction coefficient values $\mu $. Under the initial conditions (\ref{LHData}), we find that only the values $\mu <1.0131$ are admissible for the cycloid movements.
\begin{equation}
\label{CycloidTime}
T_C =\sqrt {\frac{2C_0 }{g}\left( {1+\mu ^2} \right)} 
\int\limits_0^{\tau _0 } {\sin \left( {\frac{\tau }{2}} \right)\frac{d\tau 
}{\sqrt {{\rm H}(\tau )} }} .
\end{equation}
\begin{figure}[h]
\begin{center}
\includegraphics[width=150mm]{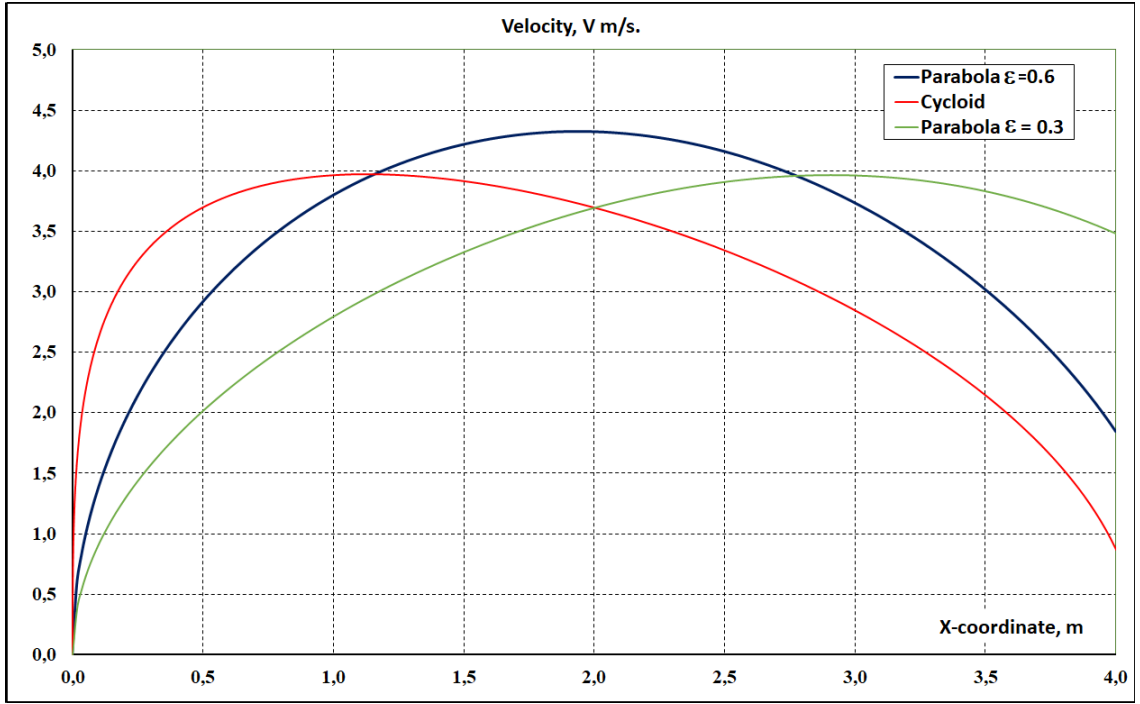}
\caption[]{Graphs of the modulus of the velocity vector on the cycloid (\ref{VelocityC}) and on parabolas (\ref{VelocityP}), expressed as functions of the horizontal coordinate $x$.}
\label{Velocity-X}
\end{center}
\end{figure}
The results of calculations using this formula are also shown in Figure \ref{Time-Parabolas3}. Comparing the data obtained, we see that in the region   $\mu <0.9$ the time of descent along the cycloid is less than along any of the parabolas mentioned above. So it is highly likely that cycloid retains the title of brachistochrone in this region. However, for large values of the friction coefficient, $\mu >0.9$, the parabola with curvature $\varepsilon =0.6$ can already lay claim to the curve of the fastest descent. The modulus of the velocity vector along the cycloid is determined by the expression:
\begin{equation}
\label{VelocityC}
\upsilon \left( \tau \right)=\frac{\sqrt {WC(\tau )} }{\sin \left( {\tau /2} 
\right)}=\sqrt {\frac{2g C_0 }{1+\mu ^2} {\rm H}\left( \tau 
\right)} .
\end{equation}
The velocities along the parabolas can be found from formulas (\ref{Wparabola}), (\ref{FParabola})
\begin{equation}
\label{VelocityP}
\upsilon \left( \alpha \right)=\frac{\sqrt {W(\alpha )} }{\cos \alpha}
=e^{\mu \alpha }\sqrt {\frac{g L}{\kappa \varepsilon } FP\left(\alpha \right)} .
\end{equation}
It is interesting to trace speed changes along the trajectory for parabola and cycloid at large values of the friction coefficient. Consider the case  $\mu =1$ when the cycloid is no longer the steepest descent curve. Figure \ref{Velocity-X} shows how the velocities are changed on each of the curves mentioned above. The parabola with curvature $\varepsilon =0.6$, provides now the smallest value for the descent time giving the following race results:
\begin{equation}
\label{TimesM1}
T_C =2.367\,s,\,\quad\,T_P \left( {\varepsilon =0.6} 
\right)=2.206\,s,\quad\,T_P \left( {\varepsilon =0.3} \right)=2.331\,s.
\end{equation}
It is remarkable that the largest speed value is reached only on this curve approximately in the middle of the trajectory. And this happens despite the fact that initial acceleration on the cycloid is the fastest. Rapid speed increase leads to the appearance of the large friction force which sharply begins to slow down the movement and ultimately leads to the fact that the point on the cycloid comes last to the finish.

\begin{figure}[h]
\begin{center}
\includegraphics[width=150mm]{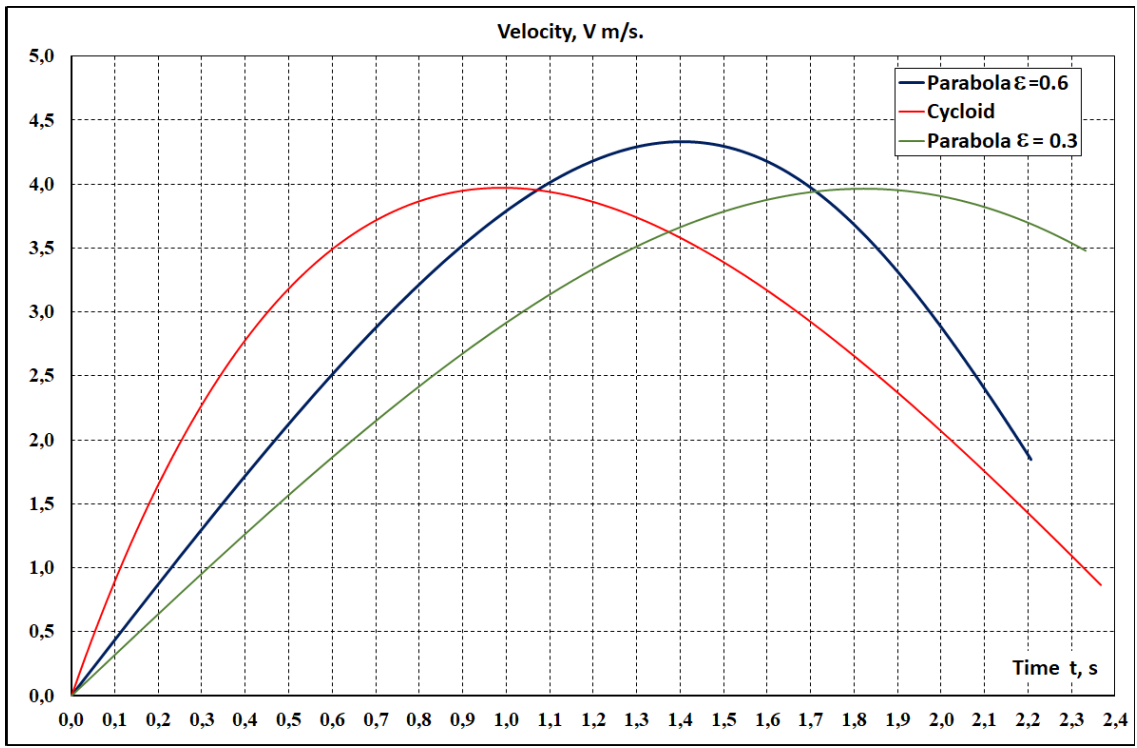}
\caption[]{ Velocity dependencies on the time.}
\label{Velocity-T}
\end{center}
\end{figure}

This observation makes it doubtful that the initial value of the slope for the true brachistochrone graph, taking into account the friction force, should tend to 90 degrees ($\alpha _0 \to \pi /2$), as is commonly believed by some authors \cite{AJM-1975,Sumbatov-2017}. Recalculating the graphs in Fig. \ref{Velocity-X} to the time dependence we arrive to results depicted in Figure \ref{Velocity-T}. As can be seen from the calculations, the parabola with the smallest descent time has an acceleration at the start about two times less than the one on the cycloid.

\begin{figure}[h]
\begin{center}
\includegraphics[width=150mm]{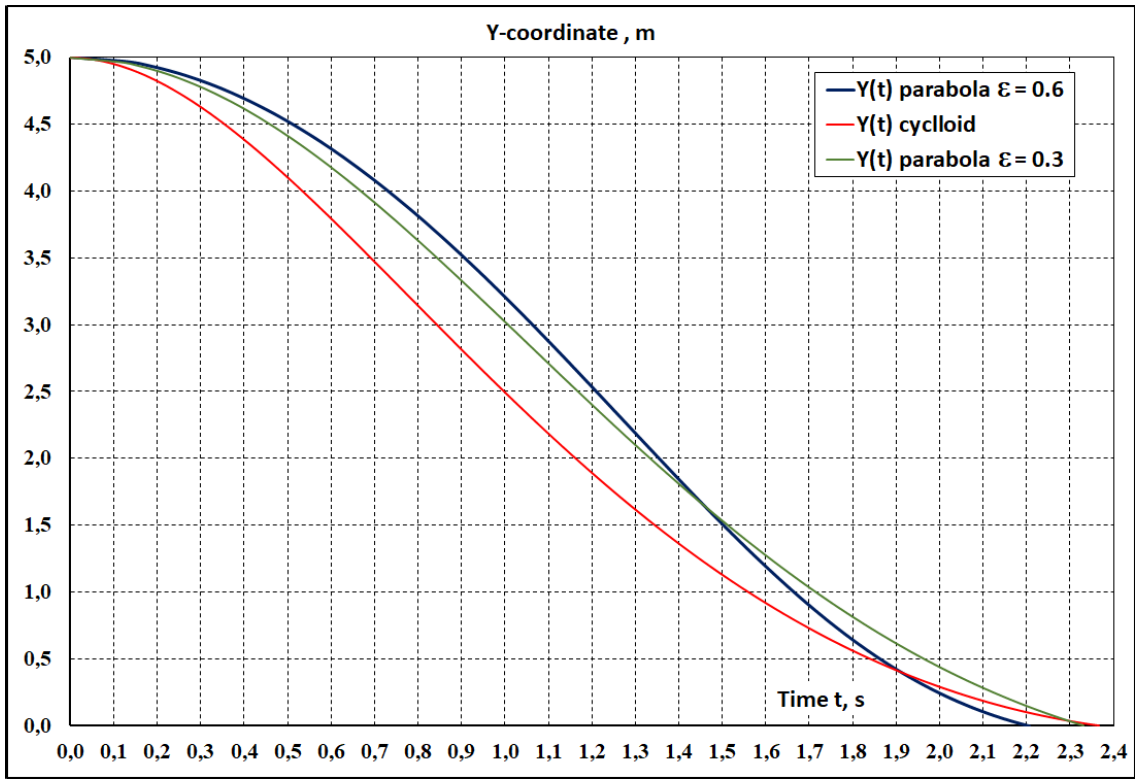}
\caption[]{Changing the vertical $Y$-coordinate of a point on a parabola and a cycloid over time.}
\label{Ycoordinate}
\end{center}
\end{figure}

\begin{figure}[h]
\begin{center}
\includegraphics[width=150mm]{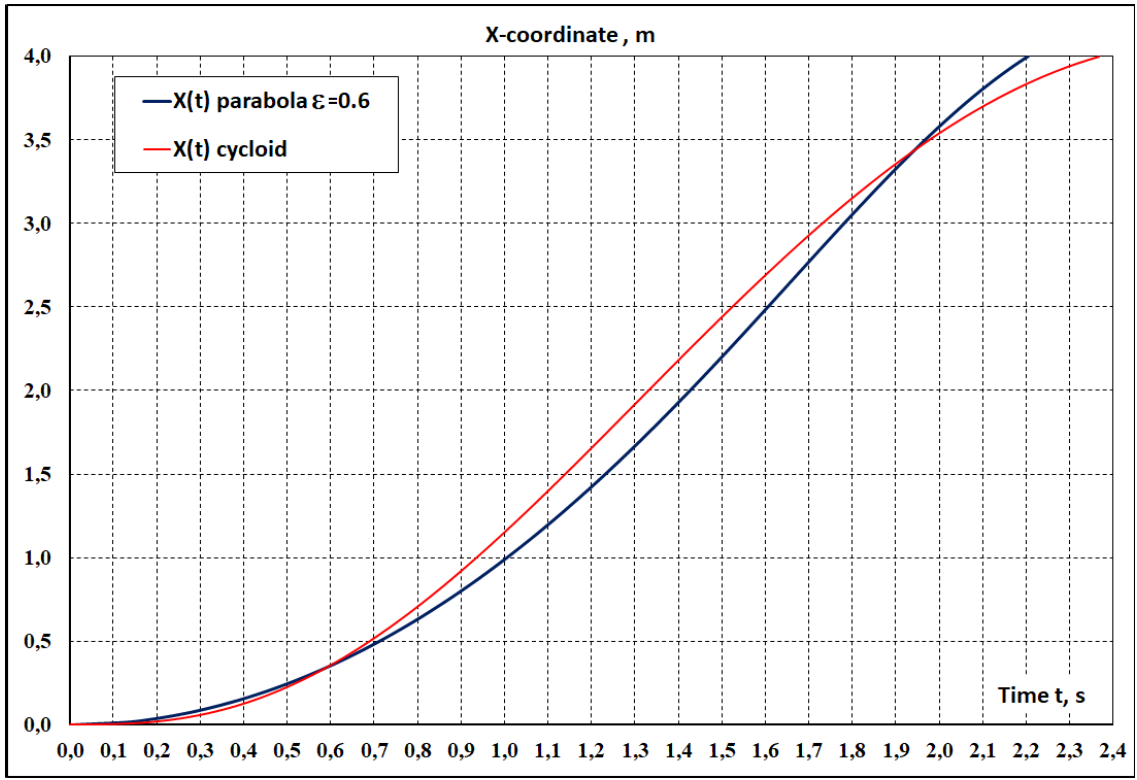}
\caption[]{Changing the horizontal $X$-coordinate of a point on a parabola and a cycloid over time.}
\label{Xcoordinate}
\end{center}
\end{figure}

Finally let us consider the laws of motion along the parabola ($\varepsilon=0.6$) and the cycloid. Figures \ref{Ycoordinate} and \ref{Xcoordinate} show how vertical and horizontal coordinates of a body on the cycloid and the parabola are changed with time. The downward movement along the cycloid is faster than the one along the parabola up to the moment $t_1 =1.9\,s.$ which is obviously due to its rapid start. However, the influence of friction forces leads to that at last part of trajectories the point on the parabola goes ahead and comes to the finish earlier.
As for the horizontal coordinate changes shown in Figure \ref{Xcoordinate}, we see that at first due to sharp steepness of the cycloid the point on it cannot overcome the one on the parabola just as for the frictionless case \cite{Kurilin-2016}. Then the body on the cycloid goes ahead at time   $t_2 =0.6\,s$ maintaining its leadership until the moment $t_3 =1.95\,s$ when the one on the parabola is again ahead of it.

\section*{Conclusions}
The proposed technique of calculations can be used for other numerical analysis of the motion dynamics with Coulomb friction forces. A wide class of the descent curves $y=\phi(x)$, can  be calculated using formulas (\ref{Weqn})-(\ref{TimeWint}). It is enough to require that the function $y=\phi(x)$  be differentiable at all points of the trajectory and then all motion parameters can be find by numerical integration. This approach is also planned to be used to calculate the fastest descent curve, brachistochrone with a given value of the sliding friction coefficient $\mu$. This is interesting both from a theoretical point of view and as a part of student teaching program with simple physical experiments. Similar physical demonstrations have been discussed by a number of different authors \cite{Phelps-1982,Deshmukh-2017,Velasco-2019,Cross-2023} and also deserve further attention and discussion.


\newpage
\begin{thebibliography}{}
\bibitem{Tikhomirov}
Tikhomirov V.M., Stories about maxima and minima, vol. 1 of Mathematical World, Universities Press, 1990, 200 p.
\bibitem{Courant}
Courant R., Robbins H., What Is Mathematics? An Elementary Approach to Ideas and Methods, Oxford University Press, 1996, 592 p.
\bibitem{Sumbatov-MIPT}
Sumbatov A.S., The problem on a brachistochrone (classification of generalizations and some recent results), 
Proceedings of Moscow Institute of Physics and Technology, 2017, {\bf 9} (3) ,66-75. (in Russian); 
\url{https://mipt.ru/upload/medialibrary/820/9_sumbatov_66_75.pdf}.
\bibitem{AJM-1975}
Ashby N., Brittin W.E., Love W.F., Wyss W., Brachistochrone with Coulomb friction, American Journal of Physics, 1975, {\bf 43}, 902-906; 
\url{https://doi.org/10.1119/1.9976}.
\bibitem{Lipp}
Lipp S.C., Brachistochrone with Coulomb friction, SIAM J. Control Optim. 1997, {\bf 35} (2), 562 –584.
\bibitem{Hayen}
Hayen J.C., Brachistochrone with Coulomb friction, International Journal of Non-Linear Mechanics, 2005, {\bf 40}, 1057–1075; 
\url{https://doi.org/10.1016/j.ijnonlinmec.2005.02.004}.
\bibitem{Golubev-2010}
Golubev Y.F., Brachistochrone with friction, Journal of Computer and Systems Sciences International, 2010,  {\bf 49}, 719–730; 
\url{https://doi.org/10.1134/S1064230710050060}.
\bibitem{Golubev-2012}
Golubev Y.F., Brachistochrone with dry and arbitrary viscous friction, Journal of Computer and Systems Sciences International, 2012, {\bf 51}, 22–37; 
\url{https://doi.org/10.1134/S1064230712010078}.
\bibitem{Sumbatov-2017}
Sumbatov A.S., Brachistochrone with Coulomb friction as the solution of an isoperimetrical variational problem, International Journal of Non-Linear Mechanics, 2017, {\bf 88}, 135–141; 
\url{https://doi.org/10.1016/j.ijnonlinmec.2016.11.002}.
\bibitem{Kurilin-2016}
Kurilin A.V., Employing the Mathcad program within the course of Theoretical Mechanics, SHS Web Conferences, 2016, {\bf 29}, 02025; 
\url{https://doi.org/10.1051/shsconf/20162902025}.
\bibitem{Phelps-1982}
Phelps F.M. III, Phelps F.M. IV, Zorn B., Gormley J., An experimental study of the brachistochrone, Eur. J. Phys. 1982, {\bf 3}, 1-4;
\url{https://doi.org/10.1088/0143-0807/3/1/001}
\bibitem{Deshmukh-2017}
Deshmukh P.C., Rajauria P., Rajans A., Vyshakh B.R., Dutta S., The Brachistochrone, 
Resonance, 2017, {\bf 22}, 847–866; 
\url{https://doi.org/10.1007/s12045-017-0539-1}.
\bibitem{Velasco-2019}
Velasco N., Vinueza D., Mármol J., Mendoza D., Pérez F., Experimental demonstration of the Brachistochrone property of the cycloid, 
J. Phys.: Conf. Ser. {\bf 1324}, 2019, 012075;
\url{https://doi.org/10.1088/1742-6596/1324/1/012075}.
\bibitem{Cross-2023}
Cross R., Sliding and rolling along circular tracks in a vertical plane, Am. J. Phys. 2023, {\bf 91}, 351;
\url{https://doi.org/10.1119/5.0107553}.







\end{thebibliography}
\end{document}